\documentclass[aps, prl, twocolumn, floatfix, superscriptaddress, reprint, 10pt]{revtex4-1}  
\usepackage{graphicx}
\usepackage[usenames]{color}
\usepackage{microtype}
\usepackage{amsmath}
\usepackage{amssymb}
\usepackage{xspace}
\usepackage{footnote}
\usepackage{relsize}
\usepackage{hhline}
\newcommand{\g}{\boldsymbol}

\newcommand{\be}{\begin{equation}}
\newcommand{\ee}{\end{equation}}

\usepackage{floatrow}
\newfloatcommand{capbtabbox}{table}[][\FBwidth]

\begin{document}

\title{
Symmetry-Adapted Machine-Learning for 
Tensorial Properties of Atomistic Systems
}

\author{Andrea Grisafi}
\affiliation{Laboratory of Computational Science and Modeling, IMX, \'Ecole Polytechnique F\'ed\'erale de Lausanne, 1015 Lausanne, Switzerland}

\author{David M. Wilkins}
\affiliation{Laboratory of Computational Science and Modeling, IMX, \'Ecole Polytechnique F\'ed\'erale de Lausanne, 1015 Lausanne, Switzerland}

\author{G\'abor Cs\'anyi}
\affiliation{Engineering Laboratory, University of Cambridge, Trumpington Street, Cambridge CB21PZ, United Kingdom}

\author{Michele Ceriotti}

\affiliation{Laboratory of Computational Science and Modeling, IMX, \'Ecole Polytechnique F\'ed\'erale de Lausanne, 1015 Lausanne, Switzerland}

\begin{abstract}
Statistical learning methods show great promise in providing an accurate prediction of materials and molecular properties, while minimizing the need for computationally demanding electronic structure calculations.
The accuracy and transferability of these models are increased significantly by encoding into the learning procedure the fundamental symmetries of rotational and permutational invariance of scalar properties.  %
However, the prediction of tensorial properties requires that the model respects the appropriate geometric transformations, rather than invariance, when the reference frame is rotated.
We introduce a formalism that can be used to perform machine-learning of tensorial properties of arbitrary rank for general  molecular geometries. To demonstrate it, we derive a tensor kernel adapted to rotational symmetry, which is the natural generalization of the smooth overlap of atomic positions (SOAP) kernel commonly used for the prediction of scalar properties at the atomic scale.
The performance and generality of the approach is demonstrated by learning the instantaneous electrical response of water oligomers of increasing complexity, from  the isolated molecule to the condensed phase.
\end{abstract}

\maketitle

The last few years have seen a surge in applications of statistical learning approaches to the prediction of the properties of molecules and materials. Chemical and materials informatics approaches -- in which large databases are mined to find correlations between structure and macroscopic properties -- have become ubiquitous~\cite{hachmann2011,hart2013,saal2013,jain2013,calderon2015,ward2017}. 
Furthermore, ``machine-learning potentials'' are increasingly used as surrogate models for  demanding electronic structure calculations, and to obtain information on the stability and properties of a material as a function of the microscopic arrangement of its atoms~\cite{behl-parr07prl,bart+10prl,mont+13njp,li+15prl,fabe+16prl}. 
For these approaches to be effective, it is crucial that the statistical learning algorithm and the mathematical representation of the atomic configurations respect the fundamental symmetries of the problem.
For example, scalar properties should be invariant under rigid translations, rotations or reflections of
the atomic configurations, as well as permutations of the order of identical atoms. 
Methods that fulfill these requirements have demonstrated very promising performance for predicting scalar quantities such as electronic ground-state energies~\cite{behl-parr07prl,bart+13prb,Rupp2015,de+16pccp,ferre2016}.

A complete description of molecular and condensed-phase systems, however, also requires the prediction of properties that are not scalars. The response of a material to mechanical, magnetic or electric  perturbations all require response coefficients that are tensorial in nature. 
The electrical response momenta -- the dipole moment $\boldsymbol{\mu}$, polarizability $\boldsymbol{\alpha}$, first hyperpolarizability $\boldsymbol{\beta}$, etc. --  underlie in particular the modelling of experiments such as infrared~\cite{Perakis2016}, Raman~\cite{Zhang2016, Heaton2008, Hutter2014, Zhang2016} and second-harmonic spectroscopy~\cite{Boyd2008,Roke2012,tocc+16jpcl}. No less importantly, they represent a fundamental ingredient to include many-body effects in atomistic simulations of a material through the development of polarizable force-fields~\cite{Morita2002,Fanourgakis2008,Cieplak2009,Lopes2009,Baker2015,Paesani2016}.

Gaussian process regression (GPR) is a commonly used machine learning technique, which is formally equivalent to kernel-ridge regression~\cite{scho+98nc,mlbook}, and is built upon the definition of a kernel function $k(\mathcal{X},\mathcal{X}')$ that encodes the similarity between two configurations $\mathcal{X}$ and $\mathcal{X}'$~ \cite{mlbook,cuturi2009,bartok2015}.
In order to guarantee that predicted properties respect the relevant physical symmetries, the kernel function must obey corresponding transformation rules. For instance, when predicting a scalar, $k(\mathcal{X},\mathcal{X}')$ should be invariant to rotations of the two configurations.
The extension to tensorial quantities is not straightforward. As discussed recently for the case of the learning of vectorial properties such as forces~\cite{glielmo}, the regression framework must be designed to that the predicted properties are covariant with respect to symmetry operations applied to the system. Under certain conditions, suitable strategies can be used to bypass the problem: for example, in the presence of relatively rigid molecular units (e.g. in water) it is possible to define a local reference frame, so that electrical response tensors can be learned  by comparing mutually aligned molecules \cite{tris+15jctc,liang2017}.
However, this approach is not generally applicable to flexible or dissociable molecular systems.  A learning algorithm that handles symmetries in a more general, mathematically rigorous fashion is required.

In this Letter, we introduce a GPR framework that explicitly includes the rotational symmetry of tensorial properties of arbitrary order, generalizing an earlier framework designed for the kernel ridge regression of forces~\cite{glielmo}, and can treat molecular or condensed-phase systems of arbitrary complexity. As a practical implementation, we define a family of kernels that are based on the smooth overlap of atomic positions (SOAP) kernels of Ref.~\cite{bart+13prb}, which we modify to account for the covariance of the tensorial property.
Within a GPR framework \cite{bartok2015}, the prediction of a property $y$ for a configuration $\mathcal{X}$ can be written as a linear combination of 
kernel functions $k(\mathcal{X},\mathcal{X}')$, that quantify the dissimilarity of the trial configuration with a set of reference inputs $\left\{\mathcal{X}_I\right\}$:
\begin{equation}
y(\mathcal{X})=\sum_I w_I k(\mathcal{X},\mathcal{X}_I).
\end{equation}
The weights can be determined by solving a linear problem $\g{w}=\left(\g{K}+\eta^2 \g{1}\right)^{-1} \g{y}$,
where $K_{IJ}=k(\mathcal{X}_I,\mathcal{X}_J)$ and $\g{y}$ contains the values of the target property for the training configurations, and $\eta$ is a regularization parameter, which can be interpreted as the expected error of the fit, due to both any intrinsic noise in the target data and the limitations of the model representation.

Consider now the case of a tensorial property $\g{T}$. We will label the components of the tensor using a compact notation $T_\mu$, where $\mu$ indicates for example a set of Cartesian axes $\mu\equiv(\alpha\beta\ldots)$.
Within a Bayesian interpretation, the kernel $k$ represents a measure of  correlations between the value of the tensorial property associated with the configurations $(\mathcal{X},\mathcal{X}')$. In particular, we can write: 
\begin{equation}\label{eq:tmu-tnu}
k_{\mu\nu}(\mathcal{X}, \mathcal{X}') = \left< T_{\mu}(\mathcal{X}) ;  T_{\nu}^\dag(\mathcal{X}') \right>,
\end{equation}
where $\left<A;B\right>$ indicates the covariance between $A$ and $B$.
In this formalism the learning algorithm is expected to simultaneously take into account all the components of~$\g{T}$.
Eq.~\eqref{eq:tmu-tnu} represents a block of a full kernel matrix, which can be built by merging the portions associated with each pair of  configurations. The complete matrix is Hermitian, so that for each block $k_{\mu\nu}(\mathcal{X}, \mathcal{X}')=k^*_{\nu\mu}(\mathcal{X}', \mathcal{X})$. 

When a generalized symmetry operation $\hat{S}$ is applied to one configuration $\mathcal{X}$ of the system, the corresponding tensorial property transforms as $T_{\mu}(\hat{S}\mathcal{X})=\sum_{\mu'}S_{\mu\mu'}T_{\mu'}(\mathcal{X})$. Then, given  two independent symmetry operations $\hat{S}$ and $\hat{S}'$ acting on the two configurations, it follows from Eqn.~\eqref{eq:tmu-tnu} that each kernel element must satisfy the following transformation rule:
\begin{equation}\label{eq:ktransform}
k_{\mu\nu}(\hat{S} \mathcal{X},\hat{S}' \mathcal{X}') = \sum_{\mu'\nu'}S_{\mu\mu'} S'_{\nu\nu'}.
k_{\mu'\nu'}(\mathcal{X}, \mathcal{X}')
\end{equation}
This is the generalization of the covariance conditions introduced in Ref.~\cite{glielmo} for the special case of learning vectors.
Similarly to that case,
one can then verify that a kernel which satisfies Eq.~\eqref{eq:ktransform} can be obtained starting from a scalar kernel $\kappa(\mathcal{X}, \mathcal{X}')$,
by averaging over the matrix that represent the symmetry operation $\hat{S}$:
\begin{equation}\label{eq:kintegral}
k_{\mu\nu}(\mathcal{X}, \mathcal{X}') = \int {\rm d}\hat{S}\ S_{\mu\nu} \kappa(\mathcal{X},\hat{S} \mathcal{X}').
 \end{equation}
The scalar kernel $\kappa$ only needs to be be independent of the absolute reference frame, but not of the relative orientation of the two configurations, i.e. $\kappa(\hat{S}\mathcal{X}, \hat{S}\mathcal{X}') = \kappa(\mathcal{X}, \mathcal{X}')$.

In the case of a Cartesian tensor $T_{\alpha\beta\ldots}$ of rank $r$, a full hierarchy of Cartesian kernels can be built by combining $r$ orthogonal rotation matrices, i.e., $S_{(\alpha\beta\ldots)(\alpha'\beta'\ldots)} = R_{\alpha\alpha'}R_{\beta\beta'}\cdots$ in Eq.\eqref{eq:kintegral}, generating a kernel with blocks of size $3^r \times 3^r$. 
However, this strategy is unnecessarily complicated.
The actual dimensionality of the problem can be significantly reduced by recasting the tensor into its irreducible spherical tensor (IST) representation $\g T:\{\g T^\lambda\}$. Each $\lambda$ identifies an orthogonal subspace of dimension $2\lambda+1$, according to $SO(3)$ algebra~\cite{Weinert1980}. Depending on the rank and the symmetries of the tensor, the decomposition contains a different number of elements, which in any case correspond to diagonal blocks of size smaller than $2r+1$. Performing a decomposition into the IST components makes the statistical learning faster and more transparent, since  each tensorial component $\g T^\lambda$ can now be independently learned as a vector of dimension $2\lambda+1$. 

What is more, in the spherical basis, the covariance conditions of Eqs.~\eqref{eq:ktransform} and \eqref{eq:kintegral} can be reformulated  by using the fact that each spherical component $\g T^\lambda$ of a completely symmetric tensor follows the same transformation rules as the corresponding vector spherical harmonics $\g Y^{\lambda}$, if co-variant, or $\g Y^{\lambda*}$, if contra-variant~\cite{Weinert1980}.
It follows that if the kernel is required to encode rotational symmetry in three dimensions, the generalized transformation matrix $S_{\mu\nu}$ of Eq.~\eqref{eq:kintegral} will be represented by the Wigner matrix $\g D^\lambda$ associated with the active rotations $\hat{R}$ of the system's configurations.~\cite{Morrison1987}

\begin{figure*}[tbhp]
 \includegraphics[width=1.0\textwidth]{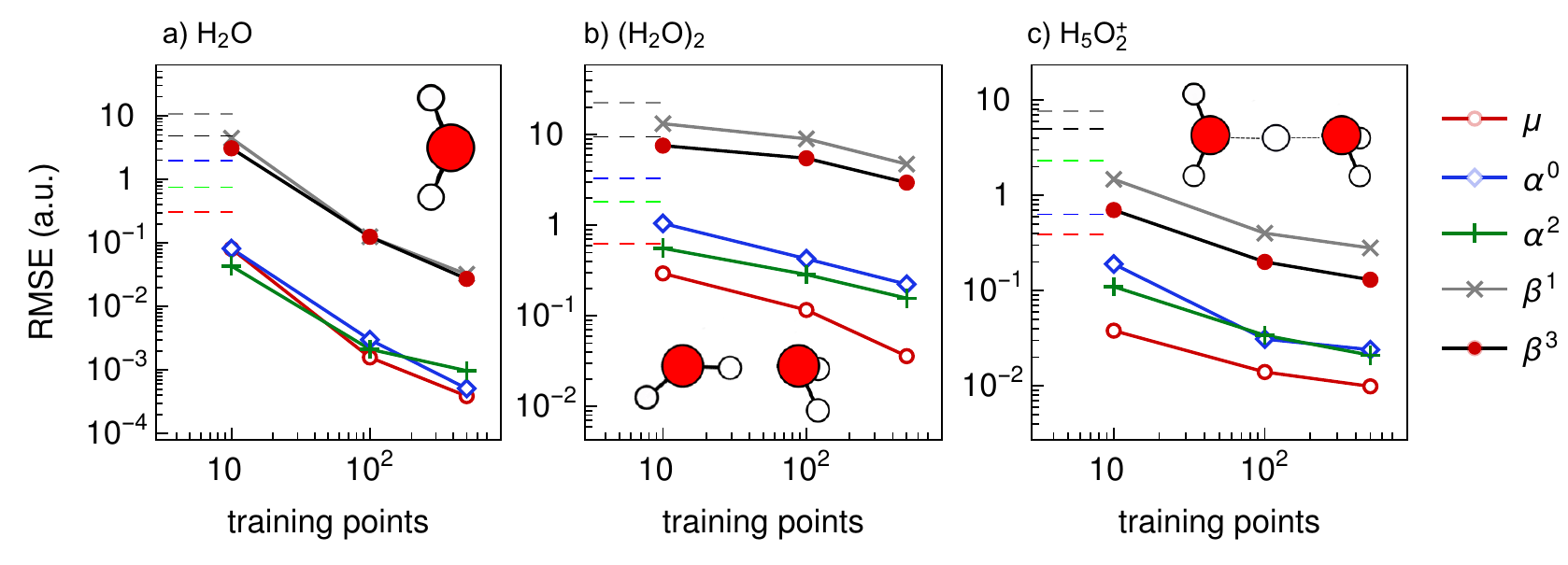}
\caption{
Learning curves of the IST components of dipole $\boldsymbol \mu$ ($\lambda=1$),  polarizability $\g \alpha$ ($\lambda=0,2$) and hyperpolarizability $\g \beta$ ($\lambda=1,3$) for water monomer (left), water dimer (center) and Zundel cation (right). For all cases the testing data set consists of 500 independent configurations. Dashed horizontal lines show the intrinsic standard deviation of the testing data set. The kernel has been computed with an environment cutoff of 4 \AA{} for the monomer and H$_5$O$_2^+$, and 5 \AA{} for the water dimer.
\label{fig:oligo}}
\end{figure*}

As a practical implementation of Eq.~\eqref{eq:kintegral}, we consider the case where $\kappa(\mathcal{X},\mathcal{X}')$ is given by the overlap between Gaussian smoothed atom densities,
\begin{equation}
\kappa(\mathcal{X},\mathcal{X}')=\left|\int \rho(\g r)\ \rho'(\g r)\,{\rm d}\g r \right|^2,
\label{scalar_k}\end{equation}
where $\rho(\g r)= \sum_{\g x\in\mathcal{X}} g_\sigma(\g r-\g x)$ is a sum over the atoms making up the environment $\mathcal{X}$ and $g_\sigma(\g r - \g x)$ is a Gaussian of width $\sigma$ centred on $\g x$.  The range of the kernel can be tuned by introducing a cutoff function that zeroes out the contribution from atom that lie farther than a given distance $r_\text{c}$ from the central atom.
With this choice of $\kappa(\mathcal{X},\mathcal{X}')$, the matrix kernel $\g k^\lambda(\mathcal{X},\mathcal{X}')$ associated with a given IST component is,
\begin{equation}\label{eq:lambda_kernel_integration}
\g k^\lambda(\mathcal{X},\mathcal{X}') = \int {\rm d}\hat R\ \g D^\lambda(\hat R)\  \left| \int \rho(\g r)\ \rho'(\hat{R} \g r)\,{\rm d}\g r\right|^2.
 \end{equation}
As shown in the Supplementary Information (SI), when an angular decomposition of the atom-centred Gaussian densities is applied~\cite{Kaufmann}, this integral can be computed analytically. The $\lambda=0$ case recovers the scalar SOAP kernel of Ref.~\cite{bart+13prb}, which has been demonstrated to be very effective for the statistical learning of scalar properties of materials and molecules~\cite{szlachta2014,Sandip,De2016,Cliffe2017,deringer2017}. 
As detailed in the SI, such a ``$\lambda$-SOAP" hierarchy of tensorial kernels can be recast as an inner product of $(2\lambda+1)$-size vectors $\g P^{\lambda}_{nn'll'}$, in the form: 
\begin{equation}\label{eq:power-spectrum}
k^{\lambda}_{\mu\nu}(\mathcal{X},\mathcal{X}') = \sum_{nn'll'}   P_{nn'll'}^{\lambda\mu}(\mathcal{X})  P_{nn'll'}^{\lambda\nu\star}(\mathcal{X}').
\end{equation}
where the contraction indexes $n,n'$ and $l,l'$ running respectively over the basis sets of the radial and the angular expansion of atomic densities.~\footnote{Note that, in agreement with the covariance condition of Eq.~\eqref{eq:ktransform}, 
the fingerprint vectors $\g P^\lambda_{nn'll'}$ transform like the $\lambda^{\rm th}$ vector spherical harmonics when the configuration $\mathcal{X}$ is rotated.}
Each $\g P^{\lambda}_{nn'll'}$ represents a symmetry-adapted fingerprint associated with the individual configurations $\mathcal{X}$, generalizing the SOAP power spectrum of Ref.~\cite{bart+13prb}.

The advantage of this formulation, which builds on a mathematically rigorous treatment of $SO(3)$ group symmetry, is that it can be applied seamlessly to molecules as well as to systems undergoing chemical reactions or to condensed phases of matter. Discrete symmetries can also be included straightforwardly. For instance, an $O(3)$ kernel with inversion symmetry can be computed as,
\begin{equation}
\g k^\lambda_{O(3)}(\mathcal{X}, \mathcal{X}') = \frac{1}{2}\left[\g k^\lambda(\mathcal{X},\mathcal{X}')+(-1)^\lambda\ \g k^\lambda(\mathcal{X},\hat{i} \mathcal{X}')\right]
\end{equation}
where $\hat{i}$ denotes inversion  and $\g k^\lambda$ is a $SO(3)$ kernel.~\footnote{In this work, the overlap between atomic densities is raised to an even power in the definition of the scalar base kernel (Eq.~\eqref{scalar_k}), which means that inversion symmetry is taken into account automatically when learning invariant properties. An overlap kernel raised to a odd power would instead be necessary when learning chiral properties~\cite{bart+13prb,de+16pccp}.}

As a demonstration of the general applicability of the framework, we now show the performance of our symmetry-adapted GPR algorithm (SA-GPR) in predicting the static polarizability series of neutral and charged water oligomers, as well as the instantaneous dielectric response tensor of liquid water configurations. Details on the training sets and the kernel hyperparameters used in each case are provided in the SI.
As a first example, we consider the polarizability series of flexible and arbitrarily-oriented water molecules in vacuum. 
The dipole moment $\boldsymbol{\mu}$, polarizability $\boldsymbol{\alpha}$ and first hyperpolarizability $\boldsymbol{\beta}$ were computed with high-end quantum chemical methods for 1000 configurations. Due to the symmetry with respect to permutations of Cartesian indices -- which is implied by the definition of response tensors as the derivatives of the electronic energy with an applied electric field -- $\boldsymbol{\alpha}$ corresponds to an irreducible representation involving the $\lambda=0$ and $\lambda=2$ spherical components only, while $\boldsymbol{\beta}$ has an IST decomposition containing $\lambda=1$ and $\lambda=3$.
Figure~\ref{fig:oligo}a shows the learning curves (i.e. the test error as a function of the number of training structures included) for all the IST components. Without explicitly providing information on the orientation of water molecules, the SA-GPR framework can easily achieve an error below 5\%{} with only 100 training points.%

A natural approach to extend the $\lambda$-SOAP framework to more complex molecules, and eventually to the condensed phase, involves decomposing the overall properties of the system into atom-centered components. 
It is straightforward to see~\cite{ourarxiv} that an atom-centered decomposition is equivalent to the learning of the system's properties using a single kernel that is built by breaking down each configuration into multiple environments, and defining the kernel of Eq.~\eqref{eq:lambda_kernel_integration} as the sum of all possible local similarities between two configurations \cite{ourarxiv},
\begin{equation}\label{eq:environments}
\g K^\lambda(\mathcal{X},\mathcal{X}') = \frac{1}{N N'}\sum_{i=1}^N\sum_{j=1}^{N'} \g k^\lambda(\mathcal{X}_{i},\mathcal{X}'_{j}),
\end{equation}
with $\mathcal{X}_{i}$ representing the $i^{\rm th}$ environment of the configuration $\mathcal{X}$.
$\g k^\lambda( \mathcal{X}_{i}, \mathcal{X}'_{j})$ is the tensorial kernel that compares the $i^{\rm th}$ local environment of the $\mathcal{X}$ configuration with the $j^{\rm th}$ local environment of the $\mathcal{X}'$ configuration.

Considering a water dimer as en example,  we take the two O atoms as centers of the environments (so that $N=2$), and
allow all of the surrounding atoms (H and O) to contribute to the smoothed atom density. The extension of this formalism to multiple chemical species involves a generalization of the scalar kernel~\eqref{scalar_k}, discussed in Ref.~\citenum{de+16pccp}.
With 500 training samples, both the isotropic and anisotropic components of  $\boldsymbol{\alpha}^{0}$ of the dimer polarizability can be learned with a RMSE below 10\%{} of the intrinsic variance (see Fig.~\ref{fig:oligo}b).
It is worth stressing that although we use the dimer responses as learning targets, the additive kernel implies a decomposition in (environment-corrected) monomer responses.  Eq.~\eqref{eq:environments} allows us to write, e.g.
\begin{equation}
\boldsymbol{\alpha}(\mathcal{X}) = \frac{1}{N} \sum_{i} \boldsymbol{\alpha}(\mathcal{X}_{i}),
\label{eq:monomer}
\end{equation}
where $\boldsymbol{\alpha}(\mathcal{X}_{i}) = \sum_{J,j} {w_{J}}N_J^{-1} \g k (\mathcal{X}_{i},\mathcal{X}^{J}_{j})$ is the contribution of the $i^{\rm th}$ environment to the dimer polarizability. 
As shown in the SI, when the two molecules are far apart the monomer polarizabilities predicted using Eq.~\eqref{eq:monomer} converge to the values computed separately for the two monomers. Thus, the discrepancy observed when the molecules separation is small can be seen as the two-body correction to the dielectric response function of individual monomers.

\begin{figure}[tbph]
    \centering
   \includegraphics[width=9cm]{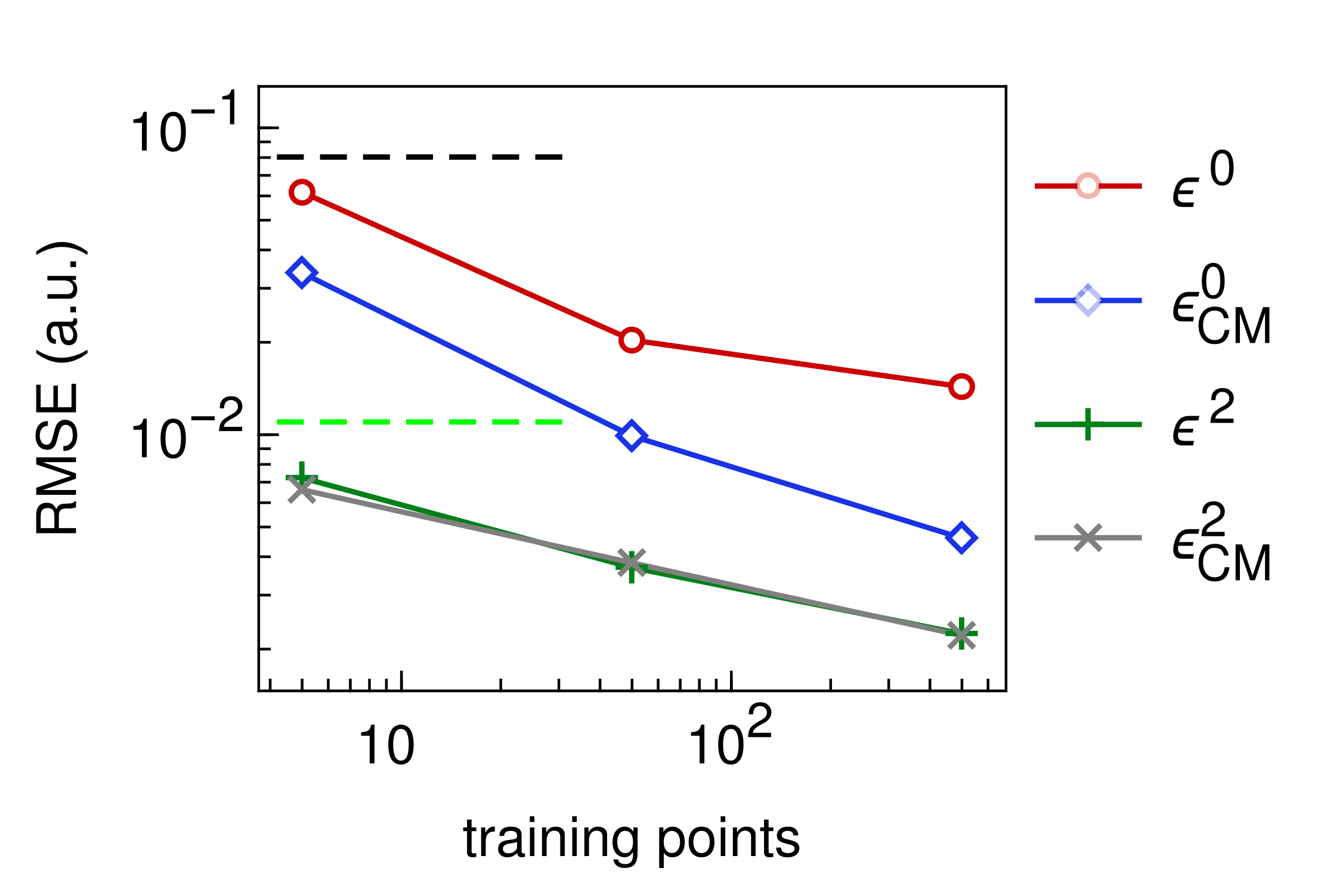}
    \caption{Learning curves of the IST components of water dielectric response tensors $\boldsymbol \varepsilon$, through direct learning (red and green lines) and indirect learning by inverting the Clausius-Mossotti (CM) relation (blue and grey lines). The testing data set consists of 500 independent configurations. Black and green dashed line refer to the intrinsic standard deviation of the testing samples for $\sigma(\g \varepsilon^0)$ and $\sigma(\g \varepsilon^2)$ respectively.}
    \label{fig:tcurve_eps0}
\end{figure}

As the next step, we consider the case of the  Zundel cation H$_{2}$O$_{5}^{+}$. 
Being both charged and chemically active, this molecule is an example of a system that would be difficult to describe in terms of separate molecular contributions. 
Fig.~\ref{fig:oligo}c compares the learning curves for the moduli of the spherical components of $\boldsymbol{\mu}$, $\boldsymbol{\alpha}$ and $\boldsymbol{\beta}$, obtained using a spherical cutoff of 4~\AA{} around each oxygen atom. Note that although each environment encompass the entire molecule, learning with atom-centered environments implies enforcing the covariance condition at the level of O atoms, which better captures the physics of the problem. The errors are well below 5\% with 500 training samples, showing that $\lambda$-SOAP kernels are well suited to extend the SA-GPR method to systems which are intrinsically not separable into smaller molecular units.

In order to test the robustness and generality of our scheme, we finally consider the prediction of the dielectric response tensor $\boldsymbol \varepsilon$ of instantaneous configurations of condensed phase water. The reference data set has been collected by computing $\boldsymbol \varepsilon$ through the modern theory of polarization~\cite{Resta2010} within density functional theory, for 400 different snapshots of a 32-molecule path integral simulation~\cite{ceri+14cpc} of room-temperature q-TIP4P/f water~\cite{habe+09jcp} (see SI for further details). 
Fig.~\ref{fig:tcurve_eps0} shows how building a $\lambda$-SOAP kernel with an environment cutoff of 4 \AA{} around each oxygen atom allows us to learn directly both the isotropic and anisotropic components of $\boldsymbol \varepsilon$ with a RMSE well below 0.01 a.u. with just 200 training samples. As we discuss in the SI, training of the isotropic component is much more effective if performed on the molecular polarizability $\boldsymbol \alpha = (\boldsymbol\varepsilon -1)(\boldsymbol\varepsilon +2)^{-1}V$. This underscores the importance of reducing the impact of non-local effects -- which appear in the definition of $\boldsymbol \varepsilon$ through the volume term -- when applying a machine-learning strategy that is based on an atom-centered decomposition.
Indeed, similar performance can be obtained by learning $\boldsymbol\varepsilon$ if $r_\text{c}$ is increased to 5 \AA{}, so that information on the volume of the simulation is captured by the kernel. 
The SA-GPR framework we introduced in this work provides a generally applicable strategy to perform kernel-based machine-learning of tensorial properties, fully incorporating their rotational symmetries. Extensions to other discrete or continuous symmetries (e.g. to cylindrical geometries, or translational invariances) is straightforward. 
Building on the existing SOAP kernel between atomic environments, we obtain a hierarchy of $\lambda$-SOAP kernels which can be used to predict the electric response tensors of systems of increasing complexity, from isolated molecules to the condensed phase. 
Being able to apply statistical learning to tensors opens the way to the prediction of anisotropic materials properties: elastic and magnetic response, NMR chemical shifts, etc. Machine-learning of molecular electric responses, which we used here as an example, makes it possible to improve the computation of linear and non-linear optical spectra, as well as to design more accurate polarizable forcefields for complex systems that cannot be described well in terms of rigid molecular entities. 
Another application with immense potential is related to the calculation of the building blocks of electronic-structure methods, such as the ground-state charge density, or the matrix elements of Hamiltonians written in an atom-centered basis. Learning the Hamiltonian would allow one to obtain ``tight-binding-like'' schemes free of an explicit parameterization, which can match the accuracy of higher levels of electronic-structure theory when computing properties such as electronic bands.
Statistical learning methods are finding applications across all fields of science and technology.
This Letter shows how to realize the full potential of these methods by making them consistent with the fundamental physical symmetries of the problem at hand.

\section*{Acknowledgments}

The authors thank Francesco Paesani for providing the water dimer structures and Giulio Imbalzano for critical reading of the manuscript. M.C was supported by the European Research Council under the European Union's Horizon 2020 research and innovation programme (grant agreement no. 677013-HBMAP). D.M.W. acknowledges funding from the Swiss National Science Foundation (Project ID 200021\_163210).

\end{document}